\documentclass[pamq,keywordsasfootnote]{ipart}

\RequirePackage{amsthm,amsmath,amssymb}
\RequirePackage{hyperref}

\startlocaldefs

\usepackage[mathscr]{eucal}

\theoremstyle{plain}
\newtheorem{thm}{Theorem}
\newtheorem{conj}{Conjecture}
\newtheorem{cor}[thm]{Corollary}
\newtheorem{prop}[thm]{Proposition}

\endlocaldefs

\pubyear{2019}
\volume{0}
\issue{0}
\firstpage{1}

\renewcommand\l{\lambda}
\newcommand\bbR{{\mathbb R}}
\newcommand\bbN{{\mathbb{N}}}
\renewcommand\S{\Sigma}

\renewcommand\d{\partial}
\newcommand\D{\nabla}
\newcommand\e{\epsilon}
\newcommand\ric{{\rm Ric}}
\newcommand\g{\gamma}
\newcommand\beq{\begin{equation}}
\newcommand\eeq{\end{equation}}
\newcommand\ben{\begin{enumerate}}
\newcommand\een{\end{enumerate}}
\newcommand\bit{\begin{itemize}}
\newcommand\eit{\end{itemize}}

\newcounter{mnotecount}

\begin{document}

\begin{frontmatter}
\title[CMC Cauchy surfaces and splitting]{Existence of CMC Cauchy surfaces \\and spacetime splitting}

\begin{aug}
    \author{\fnms{Gregory J.} \snm{Galloway}\thanksref{t2}\ead[label=e1]{galloway@math.miami.edu}}
    \address{Department of Mathematics \\ University of Miami\\ Coral Gables, FL\\
             USA\\
             \printead{e1}}
    \thankstext{t2}{Research partially supported by NSF grant DMS-1710808.}
\end{aug}
%

\begin{abstract}
In this paper, we review results on the existence (and nonexistence) of constant mean curvature spacelike hypersurfaces in the cosmological setting, and discuss the connection to the spacetime splittng problem.  It is a pleasure to dedicate this paper to Robert Bartnik, who has made fundamental contributions to this area. 
\end{abstract}

%

\end{frontmatter}

\section{Introduction}

It is a great pleasure  to contribute to this special issue  in honor of Robert Bartnik on the occasion of his 60th birthday.  I was a visiting professor in the mathematics department at UC San Diego during the 1983--1984 academic year.  That was the year S.-T. Yau arrived in the department, along with his entourage of graduate students and post docs, Robert among them.  It was an exciting year.   Robert's work has had, and continues to have, a great impact on so many people working in geometric analysis and mathematical relativity.   Personally, my interactions with Robert that year set a direction for my own research that has lasted to the present.  

Early in the year I learned from Robert, and then later from Yau, Yau's point of view concerning the Hawking-Penrose singularity theorems.  In essence, Yau's proposal  was to establish the {\it rigidity} of the  Hawking-Penrose singularity theorems.  From the standpoint of rigidity one would like to eliminate from the singularity theorems conditions like the {\it generic condition\/} and retain only weak curvature inequalities such as the {\it strong energy condition}, which requires the Ricci curvature on timelike vectors to be nonnegative.  This perspective is well-illustrated by the following `prototypical' singularity theorem, which in fact is a special case of the Hawking-Penrose singularity theorem \cite[Theorem 2, p.\ 266]{HE}.  

\newpage

\begin{thm}\label{sing} 
Let $(M,g)$ be a space-time  which satisfies the following.
\ben

\item[(a)] $M$ is globally hyperbolic with {\it compact Cauchy surfaces}.\footnote{See Section 2 for definitions.} 

\item[(b)] $M$ obeys the strong energy condition, $Ric(X,X)\ge 0$ for all timelike
vectors $X$. 

\item[(c)] $M$ satisfies the generic condition, i.e. on each inextendible timelike and null geodesic $\g$ there is a point $p$ and a vector $X \in T_pM$ orthogonal to $\g'$ such 
that  
\beq
g(R(X,\g')\g',X) \ne 0 \,,
\eeq
where $R$ is the Riemann curvature tensor.  (In other words, there is a nonzero tidal acceleration somewhere along $\g$ \cite{BEE, HE}.)
 \een
 Then $M$ is timelike or null geodesically incomplete. 
\end{thm}

Let us make a remark about the proof of this theorem.  In a spacetime with a compact Cauchy surface there is a standard procedure for constructing a causal (timelike or null) {\it line}, i.e. an inextendible causal geodesic such that each finite segment  maximizes the Lorentzian distance between its end points.  (In the case of a null line $\eta$, this is equivalent to $\eta$ being globally achronal,  meaning that no two points can be joined by a timelike curve.)  Under the assumptions of the theorem, this line cannot be complete, for otherwise the curvature conditions imply the existence of a pair of conjugate points, which would destroy the maximality.  

Note that Theorem \ref{sing} is false without the generic condition; consider, for example, the flat spacetime cylinder closed in space.  But the point of view taken here is that it should fail only under very special circumstances.  Such considerations lead to the following conjecture, explicitly stated by Bartnik as Conjecture 2 in \cite{Bart88}. 

\begin{conj}[Bartnik splitting conjecture]\label{conj}
Let $(M,g)$ be a space-time which satisfies (a) and (b) of Theorem \ref{sing}.
If $(M,g)$ is timelike geodesically complete then $(M,g)$ splits isometrically as a  product
$(\bbR\times V,-dt^2\oplus h)$, where $(V,h)$ is a compact Riemannian manifold. 
 \end{conj}

Thus, according to the conjecture,  such spacetimes $M$ must be singular (timelike geodesically incomplete) except under very special circumstances.  Remarkably, despite various attempts and partial results, the conjecture remains open in the full generality stated here.   For more recent developments, and some related results, see e.g. \cite{horo1, horo2}.  

In the next section we will review some results concerning the existence of constant mean curvature (CMC) spacelike hypersurfaces, an area to which Bartnik has made many important contributions.   
Some of the results discussed lead to partial proofs of Conjecture \ref{conj}.   We will also recall some important examples (the first due to Bartnik) of spacetimes with compact Cauchy surfaces satisfying the strong energy condition that do not admit any CMC Cauchy surfaces.   A certain CMC existence conjecture (stated in \cite{GalLing}) is also discussed.
In Section 3 we briefly comment on an entirely different approach that has been taken in an effort to prove Conjecture \ref{conj}.  This is the approach first advocated by Yau, who posed the problem \cite{Yau} of establishing a Lorentzian analogue of the Cheeger-Gromoll splitting theorem.   We also make some brief comments about Lorentzian horospheres, as developed in \cite{horo1,horo2}.   

\section{Existence and nonexistence of compact \\ CMC Cauchy surfaces}

We begin with some basic definitions and facts.   For causal theoretic notions used, but not defined below, we refer the reader to the standard references \cite{BEE, HE, ON}.  

By a spacetime we mean a time oriented Lorentzian manifold $(M,g)$ of dimension~$\ge 4$.   We assume throughout that $M$ and $g$ are smooth ($C^{\infty}$).  We further restrict attention to spacetimes $(M,g)$ that are globally hyperbolic.  Classically, this means (i) the `causal diamonds' $J^+(p) \cap J^-(q)$ are compact for all 
$p,q \in M$ and  (ii) $M$ is strongly causal.   
Following \cite{ON}, we define a Cauchy (hyper)surface in $M$ to be a subset $S$ that is met exactly once by every inextendible timelike curve.  In particular, $S$ is {\it achronal}, i.e. no two points of $S$ can be joined by a timelike curve.  In fact, it follows that a Cauchy surface $S$ is an achronal $C^0$ hypersurface that is met exactly once by every inextendible causal curve (see e.g.  \cite[Lemma 29 on p.\ 419]{ON}).  By considering the flow of a timelike vector field, one sees that any two Cauchy surfaces are homeomorphic, and that if $S$ is a Cauchy surface in $M$ then $M$ is homeomorphic to $\bbR \times S$.  
Implicit in the proof of some of the results to be discussed is the once folk theorem, and now theorem, that a smooth globally hyperbolic spacetime admits  {\it Cauchy time function} \cite{Sanchez}, i.e. a smooth time function $t$ all of whose level sets are Cauchy surfaces. 

Let $V$ be a smooth spacelike hypersurface in a spacetime $(M,g)$.  To set conventions, the second fundamental form $K$ of $V$ is defined as:  $K(X,Y) = g(\nabla_Xu,Y)$,
where $X,Y \in T_pV$, $\D$ is the Levi-Civita connection of $M$, and $u$ is the future directed timelike unit normal vector field to $V$. Then the mean curvature $H$ is given by, $H = {\rm tr}_hK$, where $h$ is the induced metric on $V$.  


To simplify certain statements, we shall use the following terminology:  A {\it CMC Cauchy surface} (resp.,  
{\it maximal Cauchy surface}), is a smooth spacelike Cauchy surface with constant mean curvature (resp., zero mean curvature).  

\subsection{Existence}

The following result relates Conjecture \ref{conj} to the existence of CMC Cauchy surfaces.  

\begin{thm}\label{maxsplit}  Let $(M,g)$ be a globally hyperbolic spacetime, which satisfies the strong energy condition (SEC), $Ric(X,X)\ge 0$ for all timelike vectors $X$.  Suppose further that $(M,g)$ contains a compact CMC Cauchy surface $S$.  If $(M,g)$ is timelike geodesically complete, then  $(M,g)$ splits isometrically as a  product $(\bbR\times S,-dt^2\oplus h)$, where $h$ is the induced metric on $S$.
\end{thm}

\noindent
This is Corollary 1 in Bartnik's paper \cite{Bart88}.  Let us make a few comments about the proof.   First of all, one can immediately reduce to the case that $S$ is maximal, $H =0$.  For if $H = c \ne 0$, then by Hawking's cosmological singularity theorem \cite[Theorem 4, p.\ 272]{HE}, $M$ would be timelike geodesically incomplete, either to the past of future.  Then, assuming $S$ is maximal,
Bartnik proves that a neighborhood $U \approx (-\e,\e) \times S$  of $S$ must be foliated by CMC Cauchy surfaces $S_t$, $t \in (-\e,\e)$.  The proof is a very nice implicit function theorem argument, which uses the linearization of the mean curvature operator acting on functions $u \in C^{\infty}(S)$ having spacelike graphs with respect  to Gaussian normal coordinates.   (It is an argument we have subsequently used in various guises over the years.)  As above, each $S_t$ must be maximal, and, in fact, must be totally geodesic, as otherwise it could be perturbed to a Cauchy surface with nonvanishing mean curvature, and again one runs into a problem with Hawking's theorem.  One is now well on the way to establishing the splitting in Theorem \ref{maxsplit}.  

Existence results for CMC Cauchy surfaces have often involved `barrier conditions',  a prime example of which is the following.   

\begin{thm}\label{cmcexist}  Let $(M,g)$ be a spacetime with compact Cauchy surfaces.   Let $S_+$ and $S_-$ be two such Cauchy surfaces, with $S_+$ in the timelike future of $S_-$, and suppose that the mean curvature $H_+$ of $S_+$ and $H_-$ of $S_-$ satisfies
\beq
H_+ < H_0 < H_-  \,,
\eeq
for some constant $H_0$.  Then there exists a Cauchy surface $S$ between $S_+$ and $S_-$ with mean curvature $H_0$. 
\end{thm} 
This theorem is a special case of the prescribed mean curvature result of Gerhardt \cite[Theorem 6.1]{Gerhardt},  and of Bartnik \cite[Theorem 4.1]{Bart84}.  (In \cite{Bart84}, Bartnik also proves, assuming a `uniform interior condition', the existence of a maximal spacelike hypersurface in the asymptotically flat setting.)  In the case that $S_-$ is future expanding, $H_- > 0$, and $S_+$ is future contracting, $H_+ < 0$, Theorem \ref{cmcexist} implies the existence of a maximal Cauchy surface between $S_+$ and $S_-$.  

Now, one would like to use Theorem \ref{maxsplit} to make some progress with Conjecture~\ref{conj}.  
Theorem \ref{cmcexist} is no help in this regard, since if $(M,g)$ satisfies the SEC, it would necessarily be timelike geodesically incomplete.  However,  after carefully studying Bartnik's thesis (see e.g.\ \cite{Bart84}, which is based on his thesis),
we saw how to give a variation of an argument in his proof of 
\cite[Theorem 4.1]{Bart84}, so as to establish the existence of a maximal Cauchy surface, subject to a certain causal theoretic condition.  

\begin{thm}[\cite{Gal84}]\label{cmcexistnooh}
Let $(M,g)$ be a spacetime with compact Cauchy surfaces, which satisfies the SEC and is timelike geodesically complete.  Suppose further that $M$ has no `observer horizons', i.e., 
\beq\label{nooh}
 I^-(\g) = I^+(\g) = M  \quad \text{for every inextendible timelike curve $\g$ in $M$.}
\eeq 
Then $M$ contains a maximal Cauchy surface.  
\end{thm} 

Existence of nontrivial observer horizons ($\d I^{\pm}(\g) \ne \emptyset$) often signal the presence of a singularity.  But, as de Sitter space shows, this need not be the case in general.  It remains open whether, for example, a future timelike geodesically complete spacetime with compact Cauchy surfaces, which satisfies the SEC, can have past observer horizons.  We come back to this point later.

\smallskip
Theorems \ref{maxsplit} and \ref{cmcexistnooh} have as an immediate consequence the following.

\begin{cor}[\cite{Gal84}]\label{split84}  
Under the additional assumption that $(M,g)$ has no observer horizons, Conjecture \ref{conj} holds.
\end{cor} 

In fact, at the time we obtained this result, Theorem \ref{maxsplit} was not available.  Instead, we were able to use a variation of Theorem \ref{maxsplit} due to Gerhardt \cite[Theorem 7.4]{Gerhardt}.  He proved, assuming the SEC, and subject to a certain compactness condition, that the region between two compact maximal hypersurfaces splits as a metric product.

We make a brief comment about the proof of Theorem \ref{cmcexistnooh}.   Bartnik's proof of the prescribed mean curvature result \cite[Theorem 4.1]{Bart84} 
is an application of  Leray-Schauder fixed point theory, employed in a suitable form.  The proof of Theorem \ref{cmcexistnooh} follows a similar approach.  One considers graphs of functions, defined on a fixed time slice $S = \{t = 0\}$, with respect to a well chosen time function $t$.  What is needed to make everything work is an a priori height estimate. Somewhat interestingly,  classical singularity theory is used to establish this.   By more or less following Bartnik's proof, we obtain a sequence of functions $u_n \in C^{\infty}(S)$ satisfying, $H(u_n)$  ($=$ mean curvature of ${\rm graph}\, u_n )= \frac1{n} u_n$.
Each such ${\rm graph}$ must meet $S$ (at which  $u_n = 0$), otherwise it would have strictly nonzero mean curvature, implying timelike geodesic incompleteness by the Hawking singularity theorem.  The no observer horizon condition  then implies (cf. \cite[Theorem 3.1]{Gal84}) that the family of graphs stays within a compact subset of $M$, hence providing the desired height estimate.   One can then use a gradient estimate and elliptic regularity to obtain a subsequence which converges in $C^{\infty}$ to a function $u$ with $H(u) = 0$. 

In \cite{Bart88}, Bartnik obtained a much stronger version of Theorem \ref{cmcexistnooh}.   
For spacetimes  $(M,g)$ with compact Cauchy surfaces, the no observer horizons condition  \eqref{nooh} was shown in \cite{Gal84} to be equivalent to several other conditions.  In particular one has the following.

\begin{prop}
Let $(M,g)$ be a spacetime with compact Cauchy surfaces.  Then 
the no observer horizons condition
\eqref{nooh} holds if and only if the sets $\d J^{\pm}(p)$ are compact for all $p\in M$.
\end{prop}
The proof makes use of basic properties of {\it achronal boundaries} \cite{Penrose}.  In particular, it is convenient to make use of the fact that the achronal boundaries $\d J^{\pm}(p)$, when compact, are Cauchy surfaces \cite[Proposition~4.8]{GalESI}.

\smallskip
Bartnik's CMC existence result in \cite{Bart88} only requires that there exists a single  point $p$ such that  
$\d J^{\pm}(p)$ are compact (and does not require timelike completeness).

\begin{thm}[\cite{Bart88}]\label{BartExist}
Let $(M,g)$ be a globally hyperbolic spacetime with compact Cauchy surfaces, which satisfies the SEC.  Suppose there exists a point $p \in M$ such that the sets  $\d J^{\pm}(p)$ are compact.
Then there exists a  CMC Cauchy surface passing through~$p$.
\end{thm} 
In fact, Bartnik stated the causality condition in this theorem in a different, but equivalent, form, namely that there exists a point $p\in M$ such that $M \setminus (I^+(p) \cup I^-(p))$ is compact.  Also, we note that the assumption of compact Cauchy surfaces  is redundant, since this follows from the assumed compactness of $\d J^{\pm}(p)$.

Bartnik's proof of Theorem \ref{BartExist} relies on his powerful result on the Dirichlet problem for the prescribed mean curvature equation,  Theorem 4.1 in \cite{BartActa}.   Let $S$ be a smooth spacelike compact Cauchy surface passing through $p$.  Consider $S_p = S \setminus \{p\}$.  The compactness of $M \setminus (I^+(p) \cup I^-(p))$, ensures that $D(S_p)$, the total domain of dependence of $S_p$, has closure contained in a compact globally hyperbolic set $K$, namely the region between two Cauchy surfaces, sufficiently far apart.   In Bartnik's terminology, $(S_p,K)$ forms a {\it standard data set}.   In the present configuration, this guarantees that for each $\lambda \in \bbR$, there exists a smooth spacelike hypersurface $V_{\l}$ in $D(S_p)$ with mean curvature $H = \l$,  such that $\d V_{\l} = \{p\}$.  (In particular,  there is no 
{\it singular set}, as defined in  \cite{BartActa}.)  The uniqueness of $V_{\l}$ is proved using the SEC, and an argument of Brill and Flaherty \cite{Brill}, so that the family $V_{\l}$, $\l \in \bbR$, forms a foliation in
$D(S_p)$.
Then by a very clever argument,  Bartnik shows that there exists some $\lambda^* \in \bbR$ such that 
$V_{\l^*}$ extends smoothly across $p$.  

\smallskip
Theorems \ref{maxsplit} and \ref{BartExist} now combine to give:

\begin{cor}[\cite{Bart88}]\label{split88}  
Under the additional assumption that there exists a point $p \in M$, such that $\d J^{\pm}(p)$ are compact, Conjecture \ref{conj} holds.
\end{cor} 

A recent paper by Dilts and Holst \cite{DH}, in which they formulate several conjectures concerning the existence of CMC Cauchy surfaces, rekindled our interest in this existence question.   Motivated by some of their considerations, Eric Ling and the author recently obtained a new CMC existence result based on a spacetime curvature condition \cite{GalLing}.   To describe this result, we recall some aspects of the {\it causal boundary} of spacetime (cf. \cite{GKP, HE}).   Heuristically, the future causal boundary $\mathscr{C}^+$ consists of `ideal points' which represent the `future end points' of future inextendible timelike curves.  
This is made precise in terms of indecomposable past sets (IPs).  Let $P$ be a past set, $P = I^-(S)$ for some set $S \subset M$.  
By definition, $P$ is an indecomposable past set if it cannot be expressed as the union of two past sets which are proper subsets of $P$.  It can be shown  \cite{BEE, HE} 
that, for strongly causal spacetimes, there are only two types of indecomposable past sets: the timelike past of a point $p$, $I^-(p)$, and the timelike past of a future inextendible timelike curve $\g$, 
$I^-(\g)$.  The latter sets are called terminal indecomposable past sets, or TIPs for short.  The future causal boundary  $\mathscr{C}^+$ is, by definition, the set of all TIPs. 
The past causal boundary $\mathscr{C}^-$ is defined in a time-dual manner.   

Note that $\mathscr{C}^+$ consists of a single point (TIP) if and only if $I^-(\g) = M$ for all future inextendible timelike curves $\g$.  A time-dual statement holds for $\mathscr{C}^+$.  Hence,  the `no observer horizon condition' \eqref{nooh} can be expressed by saying that $\mathscr{C}^+$ and $\mathscr{C}^-$ each consist of a single point.
With Bartnik's CMC existence result in mind, in \cite{Tipler}, Tipler made the following very nice observation.

\begin{prop}[Tipler \cite{Tipler}]\label{TiplerCrit}
Let $(M,g)$ be a spacetime with compact Cauchy surfaces. If $\mathscr{C}^+$ consists of a single point, then there is a point $p \in M$, sufficiently far to the future, such that $\d J^{\pm}(p)$ are compact.
\end{prop}

In fact, Tipler considers somewhat more general results, which require somewhat more involved arguments. For a simple direct proof focused on the statement above, one may consult \cite{GalLing}.   Thus, in 
Theorem \ref{BartExist} and Corollary~\ref{split88}, the condition that the sets $\d J^{\pm}(p)$ be compact for some $p$ can be replaced by the condition that the future causal boundary consists of a single point.  

We now introduce a curvature condition, related to the strong energy condition, which can be used to show under certain circumstances that the future causal boundary $\mathscr{C}^+$ consists of a single point.  Let $\Pi$ be a timelike plane in the tangent space $T_pM$.  Then the sectional curvature, $K(\Pi)$,  associated to $\Pi$ is defined as,
$$
K(\Pi) = \frac{g(R(X,Y)Y,X)}{g(X,X)g(Y,Y) -[g(X,Y)]^2} \,,
$$
where $X,Y \in T_pM$ are any two vectors spanning $\Pi$ and $R$ is the Riemann curvature tensor.  Given any unit timelike vector $u$, extend $u$ to an orthonormal basis $\{u,  e_1, ..., e_n\}$ for $T_pM$.  Then as is well-known (cf. \cite{BEE}), the spacetime Ricci curvature in the direction $u$ is given by
$$
\ric(u,u) = - \sum_{i = 1}^n K(\Pi_i)  \,,
$$
where $\Pi_i$ is the timelike plane spanned by $u$ and $e_i$.  Hence, the assumption of {\it nonpositive} timelike sectional curvatures implies the strong energy condition.  This assumption is the strongest assumption consistent with gravity being attractive.  As shown in \cite{GalLing}, assuming natural conditions on the pressure and density, perfect fluid filled FLRW spacetimes, and sufficiently small perturbations of them, have nonpositive timelike sectional curvatures.  The relevance of this curvature condition to the causal boundary was pointed out in \cite{horo2} (cf. also \cite{GalLing}).   

\begin{prop}\label{nonpos}
Let $(M, g)$ be a spacetime with compact Cauchy surfaces and with
everywhere non-positive timelike sectional curvatures, $K \le 0$.   If $(M, g )$ is future
timelike geodesically complete then the future causal boundary 
$\mathscr{C}^+$ consists of a single point.
\end{prop}

This is the time dual of  \cite[Proposition 5.11]{horo2}.  We make a comment about the proof.  Suppose there exists a future inextendible timelike geodesic $\g$ such that $I^-(\g) \ne M$.  Hence $\d I^-(\g) \ne \emptyset$.  By properties of achronal boundaries \cite{Penrose},  $\d I^-(\g)$ is an achronal $C^0$ hypersurface ruled by future inextendible null geodesics.  However, by the time-dual of Theorem~3 in \cite{EhrGal}, (see also \cite[Theorem 14.45]{BEE}), which is an application of a result of Harris (\cite[Theorem 3]{Harris} that relies on his triangle comparison theorem, any such null geodesic would eventually enter its own timelike future, contradicting the achronality of $\d I^-(\g)$.

Theorem \ref{BartExist} and Propositions \ref{TiplerCrit} and \ref{nonpos} combine to give the following.  

\begin{thm}[\cite{GalLing}]\label{CMCnonpos}
Let $(M,g)$ be a spacetime with compact Cauchy surfaces. Suppose $(M,g)$ is future timelike geodesically complete and has everywhere nonpositive timelike sectional curvatures, i.e. $K \leq 0$ everywhere. Then 
$(M,g)$ contains a CMC Cauchy surface.
\end{thm}

In particular, it follows that Conjecture \ref{conj} holds, with condition (b) replaced by the assumption of nonpositive timelike sectional curvatures.  This has also been shown to hold as an application of the Lorentzian splitting theorem \cite{EhrGal} (see the next section for further discussion).   

In the absence of an immediate counterexample, and the fact that the assumption of nonpositive timelike sectional curvatures implies the strong energy condition, one may be tempted, as indeed was done in \cite{GalLing}, to make the following conjecture.
\begin{conj}\label{conj2}
Let $(M,g)$ be a spacetime with compact Cauchy surfaces. If $(M,g)$ is future timelike geodesically complete and satisfies the SEC then $(M,g)$ contains a CMC Cauchy surface.
\end{conj}
A question related to this conjecture is whether the assumptions imply that the future causal boundary consists of a single point.  Note that a proof of this conjecture would settle Conjecture 1 in the affirmative.  
In either conjecture, if it were to help, it would be reasonable to replace ``timelike geodesically complete" by ``causal geodesically  complete".

We mention one further situation which implies the existence of a CMC Cauchy surface.

\begin{thm}\label{cmcckvf}
Let $(M,g)$ be a spacetime with compact Cauchy surfaces, which satisfies the SEC.   If $(M,g)$ admits a future complete timelike conformal Killing vector field $X$  then $(M,g)$ contains a CMC Cauchy surface.
\end{thm}

In fact, the proof of Theorem 1.3 in \cite{Costa} can be easily adapted  to show that if a spacetime $(M,g)$ with compact Cauchy surfaces admits a future complete timelike conformal Killing vector field then the future causal boundary consists of a single point.  Theorem \ref{cmcckvf} then  follows from Theorem~\ref{BartExist} and Proposition \ref{TiplerCrit}.

\subsection{Nonexistence}

In \cite{Bart88}, in addition to establishing a CMC existence result, Bartnik also put to rest the possibility that every spacetime $(M,g)$ with compact Cauchy surfaces which satisfies the SEC
 contains a CMC Cauchy surface.  We give here a rough description of his example.  Let $M_1$ be the standard future expanding dust filled FLRW model with toroidal time slices.  Let $M_2$  be the time reversed version of $M_1$, which is then past expanding.  Bartnik describes how to glue these spacetimes together across a portion of  the Schwarzschild spacetime, in a manner somewhat similar to the construction of the Openheimer-Snyder stellar model, except now each FLRW region is on the outside and the vacuum Schwarzschild region is on the inside.   The resulting spacetime $(M,g)$  satisfies the strong and weak energy conditions \cite{HE} and has compact Cauchy surfaces with topology $T^3 \# T^3$.   Bartnik gives two arguments to show that this spacetime has no CMC Cauchy surfaces, one of which is topological.   Briefly, suppose $S$ is a CMC Cauchy surface with mean curvature $H_S > 0$.   Using the time-inverting isometry that exchanges $M_1$ and $M_2$, one obtains a CMC Cauchy surface $S_*$ having mean curvature $H_{S_*} = - H(S) < 0$.  Moreover, by the Brill-Flaherty argument, $S$ is necessarily in the timelike past of $S_*$.  But,  then, by Theorem \ref{cmcexist}, there exists a maximal Cauchy surface 
$S_0 \approx T^3 \# T^3$ in between.  The Hamiltonian constraint (i.e., traced Gauss equation) and the weak energy condition would then imply that $S_0$ carries a metric of nonnegative scalar curvature. However, a manifold having topology $T^3 \# T^3$ cannot support a metric of nonnegative scalar curvature \cite{SYsc2,GL}.   Bartnik's example does not contradict Conjecture \ref{conj2}.  While it has some future complete timelike geodesics and some past complete timelike geodesics, it is neither future, nor past, timelike complete.   

In \cite{CIP}, Chru\'sciel, Isenberg and Pollack constructed an example of a maximal globally hyperbolic {\it vacuum} ($\ric = 0$) spacetime with compact Cauchy surfaces, which has no CMC Cauchy surfaces.  The example is an application of their localized gluing results for vacuum initial data sets.  They first produce a sufficiently generic vacuum initial data set 
$(g,K)$ on $V \approx T^3$.  They then perform their `connected sum' gluing procedure on the initial data sets 
$(V, g, K)$ and $(V, g, -K)$, in the process maintaining certain symmetry properties. 
The process requires that the initial data be CMC (in this case maximal) near the points where the gluing is to take place.  To accomplish this, they make use of Bartnik's results  [2] on the plateau problem for prescribed mean curvature spacelike hypersurfaces,  already discussed in connection with the proof of Theorem \ref{BartExist}.   Taking the maximal globally hyperbolic development of the resulting vacuum initial data set produces a vacuum spacetime $(M,g)$ with Cauchy surface topology $T^3 \# T^3$, and with a time-reflection property somewhat similar to that of Bartnik's example.   Similar arguments now show that 
$(M,g)$ contains no CMC Cauchy surfaces.

Until recently virtually nothing has been known about the global properties of the Chru\'sciel-Isenberg-Pollack (CIP) example (apart from the fact that it contains no CMC Cauchy surfaces).  In view of Bartnik's example and Conjecture 2, there has been an expectation that the CIP example is not  causal geodesically complete, either to the future or past.  This, in fact, has recently been shown by  Burkhart, Lesourd and Pollack. 
In \cite{Burk1}, they exploit the symmetry of the CIP construction to observe that the central 2-sphere in the connected sum is a marginally outer trapped surface, and show that this leads, for  sufficiently small ``neck size parameter"  
$\epsilon$, to future (and past) null geodesic incompleteness for the resulting spacetime.  Burkhart and Pollack \cite{Burk2} have gone on to show that this behavior is not an artifact of the symmetry but rather a consequence of the geometry of the underlying gluing construction of Isenberg, Mazzeo and Pollack \cite{IMP}, which is used in \cite{CIP}.  In particular, they show that any spacetime which arises from the IMP gluing construction must be future and past
null geodesically incomplete provided the neck size parameter is sufficiently small and the initial data set has a noncompact cover.   This is consistent with ideas associated with topological censorship, whereby one expects wormholes to pinch off to form singularities.

\smallskip
\section{Further remarks}

\subsection{Lorentzian horospheres.}  
In \cite{horo1, horo2} Carlos Vega, together with the author, developed a very general, causal theoretic based, theory of Lorentzian horospheres in globally hyperbolic spacetimes.  Let $(M,g)$  be a  globally hyperbolic future timelike geodesically complete spacetime. Very roughly speaking, a (past) Lorentzian horosphere in $(M,g)$  is obtained by taking a limit (in a precise sense) of `spheres', whose `centers' are compact sets (or a certain generalization thereof) that approach future infinity.  Lorentzian horospheres, as defined in \cite{horo1, horo2},  are always globally defined $C^0$ achronal hypersurfaces in spacetime.   

{\it Cauchy horospheres} are a particular class of horospheres defined for future timelike geodesically complete spacetimes $(M,g)$ with compact Cauchy surfaces.  To each compact Cauchy surface $S$ in 
$(M,g)$, there is an associated (past) Cauchy horosphere $S^-_{\infty}(S) \subset J^-(S)$, obtained by taking a limit of spheres of radius $k \in \bbN$, whose (sequence of) centers are the compact Cauchy surfaces 
$S_k = \{x \in M : d(S,x) = k\}$, each of which is a Lorentzian distance $k$ to the future of  $S$.
Cauchy horospheres have a number of nice geometric properties.   Assuming $(M,g)$ obeys the SEC, then a past Cauchy horosphere 
$S^-_{\infty}(S)$ is an acausal $C^0$ hypersurface with mean curvature $H \ge 0$ {\it in the support sense}.    
If, further, $(M,g)$ is past timelike geodesically complete and $S^-_{\infty}(S)$ is compact then 
$S^-_{\infty}(S)$ is a maximal Cauchy surface and $(M,g)$ splits as in Conjecture 1, with $V = S^-_{\infty}(S)$ (cf. \cite[Section 4]{horo1}).
As discussed in \cite{horo1, horo2} there are several conditions that insure, assuming timelike completeness, that $S^-_{\infty}(S)$ is compact.  These conditions will be satisfied, if, for example, the future causal boundary consist of a single point. A number of other more general rigidity results, based on horospheres, including applications with positive cosmological constant, are discussed in \cite{horo1, horo2}.  

\subsection{The Lorentzian splitting theorem}

Beyond the development of causal theory, to a large extent, the proofs of the Hawking-Penrose singularity theorems involve the extension of Riemannian comparison geometry to Lorentzian manifolds.  As an approach to the problem of establishing the rigidity of the singularity theorems, Yau posed in his well-known problem section \cite{Yau}, the problem of establishing a Lorentzian analogue of the Cheeger-Gromoll splitting theorem.   This problem was completely resolved by the end of the 80's.  The most basic version of the Lorentzian splitting theorem, due to Eschenburg \cite{Esch}, is as follows.

\begin{thm} [Lorentzian Splitting Theorem] \label{LST}
Let $M$ be a globally hyperbolic, timelike geodesically complete spacetime, satisfying the strong energy condition, $\mathrm{Ric}(X,X) \ge 0$, for all timelike $X$. If $M$ admits a timelike line, then $M$ splits as an isometric product, 
$(M^{n+1}, g) \approx (\bbR \times \S^n, -dt^2 \oplus h)$  where $\S^n$ is a smooth, geodesically complete, spacelike hypersurface, with induced metric $h$.
\end{thm}

In \cite{GalLST} and \cite{Newman} proofs were obtained under weaker assumptions.  In \cite{GalLST}, we were able to prove the above by only assuming that the timelike line is complete, while retaining the global hyperbolicity assumption.  In \cite{Newman}, Newman gave a proof assuming timelike geodesic completeness, but without assuming global hyperbolicity, which corresponds to the original version posed by Yau. All of the proofs are based on an analysis of the Lorentzian Busemann functions $b^{\pm}_{\g}$ associated to the given 
timelike line $\g$.  In our proof, we established a maximum principle for Lorentzian Busemann functions restricted to a {\it maximal} spacelike hypersurface. We then used this to show that $b^{\pm}_{\g}$ agree along a maximal hypersurface $S$ defined near $\g(0)$.  One is then well on the way to establishing a splitting in a neighborhood of $\g$. The proof is completed by globalizing as in \cite{Esch}.   Where does one go for the maximal hypersurface?  As in previous examples discussed above, one appeals to Bartnik's basic existence result \cite{Bart84} on the Dirichlet problem for the prescribed mean curvature equation, which, as was needed in the present case, allows for rough boundary data.  This technique was again used in \cite{Newman}, and in general it simplifies a number of arguments in the proof of the Lorentzian splitting theorem. 

One would like to use Theorem \ref{LST} to prove Conjecture \ref{conj}.  As mentioned in the introduction, there is a standard procedure to construct a causal line, i.e. a causal geodesic each segment of which maximizes the Lorentzian distance, in a spacetime $(M,g)$ with a compact Cauchy surface $S$:  One takes a sequence of points $\{q_n\} \subset I^+(S)$ extending to future infinity, and a sequence of points $\{p_n\} \subset I^-(S)$, with $p_n \in I^-(q_n)$,  extending to past infinity, and considers the maximal timelike geodesic segment $\g_n$ from $p_n$ to $q_n$, each of which must intersect $S$.  By the compactness of $S$, some subsquence will converge to a causal (timelike or null) line $\g$.  Unfortunately, without any further assumptions, this limit line $\g$ need not be timelike; the segments $\g_n$ can turn null in the limit.  
An unpleasant example is given in \cite{EhrGal} of a causal geodesically complete spacetime with compact Cauchy surfaces, which does not contain any timelike lines.  However, this example does not satisfy any nice curvature conditions.  

It is easy to see that the condition in Corollary \ref{split88}, that the sets $\d J^{\pm}(p)$ be compact for some $p \in M$, is sufficient to rule out null lines.  Indeed, any inextendible null geodesic $\eta$ must meet  both $\d J^-(p)$ and $\d J^+(p)$ (as they are Cauchy surfaces), from which it can easily be argued that $\eta$ is not achronal. In this case $(M,g)$ contains a timelike line, and Theorem \ref{LST} can be used to obtain Corollary \ref{split88}.  As discussed and reviewed in \cite{horo1, horo2}, there are several other conditions that ensure the existence of a timelike line, and hence, under which Conjecture \ref{conj} holds.  
All such conditions are implied by the condition that the future causal boundary consists of a single point.  Nevertheless, as mentioned at the beginning, Conjecture 1 remains open in  full generality.   In consideration of these issues, in \cite{Gnullmp} we studied the rigidity of spacetimes satisfying the null energy condition and containing a null line.  There we showed, under the assumption of null geodesic completeness, that a null line must be contained in an embedded achronal totally geodesic null hypersurface; in effect, another CMC result.

\section*{Acknowledgements}

We would like to thank Piotr Chru\'sciel, Ivan Costa e Silva, Eric Ling, Dan Pollack and Carlos Vega for helpful comments during the preparation of this paper.


\bibliographystyle{imsart-number}
\bibliography{bart}

\end{document}